\documentclass[12pt]{article}

\usepackage{amsmath}

\renewcommand{\theequation}{\arabic{equation}}
\newcommand{\EQ}{\begin{equation}}
\newcommand{\EN}{\end{equation}}

\newcommand{\ket}[1]{\left|#1\right\rangle}      

\newcommand{\IM}{\mathbf{\imath}}

\newcommand{\bear}{\begin{eqnarray}}
\newcommand{\ear}{\end{eqnarray}}
\newcommand{\bt} { \begin{tabular} }
\newcommand{\et}{ \end{tabular} }
\newcommand{\bc} { \begin{center} }
\newcommand{\ec}{ \end{center} }

\newcommand{\btb} { \begin{table} }
\newcommand{\etb}{ \end{table} }

\begin{document}

\topmargin 0pt
\oddsidemargin 5mm
\newcommand{\NP}[1]{Nucl.\ Phys.\ {\bf #1}}
\newcommand{\PL}[1]{Phys.\ Lett.\ {\bf #1}}
\newcommand{\NC}[1]{Nuovo Cimento {\bf #1}}
\newcommand{\CMP}[1]{Comm.\ Math.\ Phys.\ {\bf #1}}
\newcommand{\PR}[1]{Phys.\ Rev.\ {\bf #1}}
\newcommand{\PRL}[1]{Phys.\ Rev.\ Lett.\ {\bf #1}}
\newcommand{\MPL}[1]{Mod.\ Phys.\ Lett.\ {\bf #1}}
\newcommand{\JETP}[1]{Sov.\ Phys.\ JETP {\bf #1}}
\newcommand{\TMP}[1]{Teor.\ Mat.\ Fiz.\ {\bf #1}}

\renewcommand{\thefootnote}{\fnsymbol{footnote}}

\newpage
\setcounter{page}{0}
\begin{titlepage}
\begin{flushright}
UFSCARF-TH-09-21
\end{flushright}
\vspace{0.5cm}
\begin{center}
{\large The spectrum of an open vertex model based } \\
{\large on the  $U_q[SU(2)]$
algebra at roots of unity }\\
\vspace{1cm}
{\large M.J. Martins} \\
\vspace{0.15cm}
{\em Universidade Federal de S\~ao Carlos\\
Departamento de F\'{\i}sica \\
C.P. 676, 13565-905, S\~ao Carlos(SP), Brazil}\\
\vspace{0.35cm}
{\large C.S. Melo} \\
\vspace{0.15cm}
{\em Universidade de S\~ao Paulo\\
Instituto  de F\'{\i}sica \\
05315-970, S\~ao Paulo(SP), Brazil}\\
\end{center}
\vspace{0.5cm}

\begin{abstract}
We study the exact solution of an $N$-state vertex model based on the
representation of the $U_q[SU(2)]$ algebra at roots of unity with
diagonal open boundaries. We find that the respective reflection
equation provides us one general class of diagonal $K$-matrices
having one free-parameter. We determine the eigenvalues of the
double-row transfer matrix and the respective Bethe
ansatz equation within the algebraic Bethe ansatz
framework. The structure of the Bethe ansatz equation
combine a pseudomomenta function depending on a free-parameter
with
scattering phase-shifts that are fixed by the
roots of unity and boundary variables.

\end{abstract}

\vspace{.15cm} \centerline{PACS numbers:  05.50+q, 02.30.IK}
\vspace{.1cm} \centerline{Keywords: Lattice
Integrable Models, Open Boundary Conditions, Bethe Ansatz}
\vspace{.15cm} \centerline{November
2009}

\end{titlepage}


\pagestyle{empty}

\newpage

\pagestyle{plain}
\pagenumbering{arabic}

\renewcommand{\thefootnote}{\arabic{footnote}}

\section{Introduction}

A relevant class of two-dimensional integrable models are those
based on the representations of a given quantum affine algebra
$U_q[G]$ \cite{QG}. The quantum group framework permits us to
generate solutions of the Yang-Baxter equation which can be seen as
the Boltzmann weights of two-dimensional vertex models \cite{BA}.
The standard paradigm is the spin-$s$  representation of the
$U_q[SU(2)]$ algebra for generic values of $q$ \cite{QG1} which
leads us to an $N$-state extension of the six-vertex model
\cite{SO,FA}. The corresponding one-dimensional spin chain turns out
to be an integrable generalization of the Heisenberg model with
arbitrary spin $s=(N-1)/2$ \cite{RES}.

The previous example by no means exhaust the possibility of
constructing vertex models within the quantum $U_q[SU(2)]$ algebra.
It still remains the representations for non-generic values of $q$
which are known to be quite different from the one with arbitrary
$q$ \cite{QG2}. In fact, a new family of $N$-state vertex model can
be generated by exploring non-cyclic $U_q[SU(2)]$ representations
when $q$ is a root of unity. The respective $R$-matrix depends on
both an extra continuous variable besides the spectral parameter and
on a discrete variable characterizing the possible roots of unity
branches. We remark that this $R$-matrix has been previously
discussed in the literature in distinct contexts such as on the realm
of new braid matrices with extra color variables \cite{DEG}, in the
Baxterization of braids for the special cases of $N=2,3$ 
\cite{CO} and in the quantum group framework \cite{SIE,DEG1}.

The Bethe ansatz solution of such $N$-vertex model based on roots of
unity has so far been restricted to the case of periodic boundary
conditions \cite{SIE1,MC}. Results for fixed boundary conditions are
concentrated on the calculations of the partition function with
certain domain wall boundary conditions \cite{CARA}. The purpose of this paper
is to start the
study of this class of vertex models with open integrable boundary
terms. We shall here consider the simplest case of diagonal boundary
conditions containing a free parameter. We find that the dependence
of the Bethe equations on the continuous parameter characterizing
the roots of unity representation appears only in the pseudomomenta
function. By way of contrast the respective many-body phase-shifts are functions
determined by the discrete roots of unity branches and the boundary parameters.

The outline of this paper is as follows. In section 2 we review the
structure of the bulk $R$-matrix and discuss properties that are
useful to build up boundary $K$-matrices. In section 3 we discuss
the most general diagonal solution of the reflection equations for
both left and right boundaries. We also present the explicit
expressions of the corresponding one-dimensional open spin chains
for $N=2$ and $N=3$. In section 4 we adapt the algebraic Bethe
ansatz construction of \cite{CGM} in order to derive the eigenvalues
of the double-row transfer matrix and the respective Bethe
equations. Our conclusions are summarized in section 5. In
Appendices A and B we present technical details concerning the form of
projectors and $R$-matrix amplitudes.

\section{The $R$-matrix properties}

The corresponding $R$-matrix of the $N$-state vertex model based on
the $U_q[SU(2)]$ algebra at roots of unity has been previously
discussed by several authors \cite{DEG,CO,SIE,DEG1}. We shall follow
here the recent presentation given by us in terms of the
Baxterization approach \cite{MC}. For practical computations it is
convenient to write the $R$-matrix introducing an auxiliary operator
$\check{R}(\lambda)$,
\EQ
R_{12}(\lambda)=P_{12}
\check{R}_{12}(\lambda) , \label{eqt1}
\EN
where $P_{12}$  denotes the
$C^{N} \otimes C^{N}$ operator.

The matrix $\check{R}(\lambda)$ can be expressed by means of the
following linear combination of projectors,
\EQ
\check{R}(\lambda) =
\rho(\lambda) \sum_{i=1}^{N} \prod_{j=i}^{N-1} \frac{\sinh[\frac{\IM
\pi k (j-1)}{N}+\IM \gamma+\lambda]}{\sinh[\frac{\IM \pi k
(j-1)}{N}+\IM \gamma-\lambda]} P_i(\gamma,k) ,
 \label{eqt2}
\EN
where $k$ is an integer coprime to $N$ and the overall normalization $\rho(\lambda)$ is chosen,
\EQ
\rho(\lambda)=\prod_{j=1}^{N-1}\sinh[-\lambda+\IM \gamma+\frac{\IM \pi k
(j-1)}{N}].
\label{eqt3}
\EN

The $R$-matrix (\ref{eqt1}-\ref{eqt3}) is characterized by the
continuous variable $\gamma$ and the discrete index $k$
parameterizing the non-cyclic $U_q[SU(2)]$ representation at the
roots of unity $q=e^{\frac{2 \IM \pi k}{N}}$. The projectors
$P_i(\lambda,k)$ can be written in terms of the corresponding braid
representation \cite{DEG} whose expressions can be found in
\cite{MC}. To make this paper self-consistent we have summarized the
main formula for $P_i(\gamma,k)$ in Appendix A.
In order to investigate integrable open boundary conditions it is
convenient to bring the bulk $R$-matrix (\ref{eqt1}-\ref{eqt3}) to
its most possible symmetrical form. As usual this is accomplished by
performing a transformation in $R$-matrix (\ref{eqt1}-\ref{eqt3})
that preserves the Yang-Baxter equation. We find that the suitable
spectral parameter dependent transformation is,
\EQ
\bar{R}_{12}(\lambda)=V_1(\lambda)
{R}_{12}(\lambda)V_1^{-1}(\lambda).
\EN
where the gauge matrix $V(\lambda)$ is diagonal and it is given by
\EQ
V(\lambda)= \sum_{a=1}^{N} e^{\lambda (a-1)} e_{a, a}
\label{gaugem}
\EN
while
$e_{a, b}$ denotes standard $N \times N$ Weyl matrices.

Let us now discuss the structure of the $R$-matrix
$\bar{R}_{12}(\lambda)$ for few values of $N$. The simplest case
$N=2$ turns out to be directly related to a six-vertex model
satisfying the free-fermion condition, namely \bear
\bar{R}_{12}(\lambda)=
\left(\begin{array}{cc cc} \sinh[\IM \gamma+\lambda] & 0 & 0 & 0 \\
0 & \sinh[\lambda] & \sinh[\IM \gamma] & 0 \\
0 & \sinh[\IM \gamma] & \sinh[\lambda] & 0 \\
0 & 0 & 0 & \sinh[\IM \gamma-\lambda]
\end{array}\right) \label{RweightS1p2}
\ear

The $R$-matrix (\ref{eqt1}-\ref{eqt3}) for $N \ge 3$ gives origin
to novel integrable vertex models. For the simplest
case $N=3$ one has a nineteen-vertex model and the structure of the
corresponding $R$-matrix is,
\bear
\bar{R}_{12}(\lambda)= \left(\begin{array}{ccc ccc ccc}
a_+(\lambda) & 0 & 0 & 0 & 0 & 0 & 0 & 0 & 0 \\
0 & b_+(\lambda) & 0 & c_+(\lambda) & 0 & 0 & 0 & 0 & 0 \\
0 & 0 & f(\lambda) & 0 & d(\lambda) & 0 & e(\lambda) & 0 & 0 \\
0 & c_+(\lambda) & 0 & b_+(\lambda) & 0 & 0 & 0 & 0 & 0 \\
0 & 0 & d(\lambda) & 0 & g(\lambda) & 0 & d(\lambda) & 0 & 0 \\
0 & 0 & 0 & 0 & 0 & b_-(\lambda) & 0 & c_-(\lambda) & 0 \\
0 & 0 & e(\lambda) & 0 & d(\lambda) & 0 & f(\lambda) & 0 & 0 \\
0 & 0 & 0 & 0 & 0 & c_-(\lambda) & 0 & b_-(\lambda) & 0 \\
0 & 0 & 0 & 0 & 0 & 0 & 0 & 0 & a_-(\lambda) \\
\end{array} \right)
\ear

The
respective Boltzmann weights are given by,
\bear
a_{\pm}(\lambda) &=& \sinh[\IM \gamma \pm \lambda] \sinh[\IM \gamma +
\frac{\IM \pi k}{3} \pm \lambda]
\nonumber \\
b_{+}(\lambda) &=& \sinh[\lambda] \sinh[\IM \gamma + \frac{\IM \pi k}{3} +
\lambda],~~~~ b_{-}(\lambda)=\varepsilon_k \sinh[\lambda] \sinh[\IM \gamma -
\lambda]
\nonumber \\
c_{+}(\lambda) &=& \sinh[\IM \gamma] \sinh[\IM \gamma + \frac{\IM \pi k}{3} +
\lambda]
,~~~~
c_{-}(\lambda)=\sinh[\IM \gamma - \lambda] \sinh[\IM \gamma + \frac{\IM \pi
k}{3}]
\nonumber \\
d(\lambda)  &=&  \varepsilon_k^{3/2} \sinh[\lambda] \sqrt{
\sinh[\IM \gamma] \sinh[\IM \gamma + \frac{\IM \pi k}{3} ] }
,~~~~
e(\lambda) = \sinh[\IM \gamma] \sinh[\IM \gamma + \frac{\IM \pi k}{3}]
\nonumber \\
f(\lambda) &=& \sinh[\lambda] \sinh[\lambda + \frac{\IM \pi k}{3}]
,~~~~
g(\lambda) = \sinh[\IM \gamma] \sinh[\IM \gamma + \frac{\IM \pi
k}{3}]+\sinh[\lambda - \frac{\IM \pi k}{3}] \sinh[\lambda]
\label{RweightS1}
\ear
where the phase $\varepsilon_k= \exp[\IM \pi (k-1)]$.

Considering Eqs.(\ref{eqt1}-\ref{gaugem}) and the 
expressions for the projectors (\ref{proj1}-\ref{proj2}) 
one is able to compute the $R$-matrix elements of
$\bar{R}_{12}(\lambda)$ on the Weyl basis,
\EQ
\bar{R}_{12}(\lambda)
= \sum_{a,b,c,d=1}^{N} \bar{R}_{a, b}^{c, d}(\lambda) e_{a, c}
\otimes e_{b, d} \label{Rmatrix}
\EN
for rather high values of $N$
with moderate computational effort. As further examples we have
exhibited in Appendix B the non-null Boltzmann weights of the
$\bar{R}_{12}(\lambda)$
for $N=4$ and $N=5$.

We now turn our attention to describing the symmetry properties of
$\bar{R}_{12}(\lambda)$. Besides satisfying the unitarity property
it is invariant by both temporal and parity symmetry, namely
\bear
\bar{R}_{12}(\lambda) \bar{R}_{21}(-\lambda) = \rho(\lambda)
\rho(-\lambda) I_{N} \otimes I_{N} \label{unitarity} \\
P_{12} \bar{R}_{12}(\lambda) P_{12} = \bar{R}_{12}(\lambda) \label{permutation} \\
\bar{R}_{12}(\lambda)^{t_1 t_2} = \bar{R}_{12}(\lambda)
\label{transposition} \ear
where $I_N$ is the $N \times N$ identity
matrix and $t_i$ denotes the transposition on the $i$-th space.

We also note that for arbitrary $\gamma$ the matrix elements
$\bar{R}_{a, b}^{c, d}(\lambda)$ of $\bar{R}_{12}(\lambda)$ are not
self-conjugated under the charge invariance\footnote{Recall that the
charge symmetry relates the weights $\bar{R}_{a, b}^{c, d}(\lambda)$
and $\bar{R}_{N+1-a,N+1-b}^{N+1-c,N+1-d}(\lambda)$.}. As a
consequence of that the $\bar{R}_{12}(\lambda)$ does not satisfy
standard crossing properties but only an analog of that which is given by,
\EQ
\bar{R}_{12}(\lambda)^{t_2} \bar{R}_{21}(-\lambda+\frac{2 \pi \IM
k}{N})^{t_2} = \rho_1(\lambda) I_{N} \otimes I_{N}
\label{crossing}
\EN
where
$ \rho_1(\lambda) = \prod_{j=1}^{N-1}
\sinh^2[\lambda+\frac{\IM \pi k (j-1)}{N}].
$

In next section we shall use the properties
(\ref{unitarity}-\ref{crossing}) to obtain the right $K$-matrix
$K^{(+)}(\lambda)$ from the left one $K^{(-)}(\lambda)$ by means of
an isomorphism.

\section{The $K$-matrices and open spin chains}

The integrability at the boundary is described in terms of two
scaterring matrices $K^{(\pm)}(\lambda)$ \cite{SK}. The
compatibility between bulk and boundary scattering lead us to an
algebraic condition at one of the ends of an open chain, which reads
\cite{SK}
\EQ
\bar{R}_{12}(\lambda-\mu) K_1^{(-)}(\lambda)
\bar{R}_{21}(\lambda+\mu) K_2^{(-)}(\mu) =
 K_2^{(-)}(\mu) \bar{R}_{12}(\lambda+\mu)
 K_1^{(-)}(\lambda) \bar{R}_{21}(\lambda-\mu) \label{reflection}
\EN
where $ K_1^{(-)}(\lambda) = K^{(-)}(\lambda) \otimes I_{N}$ and
$ K_2^{(-)}(\lambda) = I_{N} \otimes  K^{(-)}(\lambda)$.

When the bulk $R$-matrix satisfies the properties
(\ref{unitarity}-\ref{crossing}) one can follow a procedure devised
in \cite{NEP1,NEP2} to obtain the matrix $K^{(+)}(\lambda)$ at the
opposite boundary. At this point we note that property
(\ref{crossing}) plays the role of the crossing symmetry \cite{NEP2}
and the matrix $K^{(+)}(\lambda)$ is then fixed by the following
isomorphism,
\EQ K^{(+)}(\lambda) = K^{(-)} (\frac{\IM \pi
k}{N}-\lambda)^t.
\label{isomorphism}
\EN

As a consequence of that we are left with the task to solve a single
reflection equation. Here we shall be searching only for diagonal
solutions of the reflection equation (\ref{reflection}), namely
\EQ
K^{(-)}(\lambda)= \sum_{a=1}^{N} K_a^{(-)}(\lambda) e_{a, a}.
\label{Kmatrix}
\EN

The next step is to substitute the ansatz (\ref{Kmatrix}) in the
reflection equation (\ref{reflection}) and look for the simplest
relations constraining the unknown matrix elements
$K_a^{(-)}(\lambda)$. Among the many functional equations we find
that there exists two of them that fix the ratios between the first
two and the last two elements in a rather suitable way,
\EQ
\frac{K_2^{(-)}(\mu)}{K_1^{(-)}(\mu)} =
\frac{\frac{K_2^{(-)}(\lambda)}{K_1^{(-)}(\lambda)}
\bar{R}_{1,2}^{2,1}(\lambda-\mu) \bar{R}_{2,1}^{2,1}(\mu+\lambda) +
\bar{R}_{2,1}^{2,1}(\lambda-\mu) \bar{R}_{2,1}^{1,2}(\mu+\lambda) }{
\bar{R}_{2,1}^{1,2}(\lambda-\mu) \bar{R}_{2,1}^{2,1}(\mu+\lambda) +
\frac{K_2^{(-)}(\lambda)}{K_1^{(-)}(\lambda)}
\bar{R}_{2,1}^{2,1}(\lambda-\mu) \bar{R}_{1,2}^{2,1}(\mu+\lambda) }
\label{functional1}
\EN
and
\EQ
\frac{K_{N}^{(-)}(\mu)}{K_{N-1}^{(-)}(\mu)} =
\frac{\frac{K_{N}^{(-)}(\lambda)}{K_{N-1}^{(-)}(\lambda)}
\bar{R}_{N-1,N}^{N-1,N}(\lambda+\mu) \bar{R}_{N-1,N}^{N,N-1}(\lambda-\mu) +
\bar{R}_{N,N-1}^{N-1,N}(\lambda+\mu) \bar{R}_{N-1,N}^{N-1,N}(\lambda-\mu) }{
\bar{R}_{N-1,N}^{N-1,N}(\lambda+\mu) \bar{R}_{N,N-1}^{N-1,N}(\lambda-\mu) +
\frac{K_{N}^{(-)}(\lambda)}{K_{N-1}^{(-)}(\lambda)}
\bar{R}_{N-1,N}^{N,N-1}(\lambda+\mu) \bar{R}_{N-1,N}^{N-1,N}(\lambda-\mu) }.
\label{functional2}
\EN

Taking into account the explicit expressions for the Boltzmann
weights we conclude that the right-hand sides of
(\ref{functional1},\ref{functional2}) are independent of variable $\lambda$
only if the ratios $\frac{K_2^{(-)}(\lambda)}{K_1^{(-)}(\lambda)}$
and $\frac{K_{N}^{(-)}(\lambda)}{K_{N-1}^{(-)}(\lambda)}$ satisfy
the following relations
\EQ
\frac{K_{2}^{(-)}(\lambda)}{K_{1}^{(-)}(\lambda)} =
\frac{\sinh[\beta_1-\lambda]}{\sinh[\beta_1+\lambda]}, ~~
\frac{K_{N}^{(-)}(\lambda)}{K_{N-1}^{(-)}(\lambda)} =
\frac{\sinh[\beta_{N-1}-\lambda]}{\sinh[\beta_{N-1}+\lambda]},
\label{Ksol} \EN
where $\beta_1$ and $\beta_{N-1}$ are free
continuous parameters.

In the particular cases $N=2$ and $N=3$ the previous analysis
already provides us a proposal for the $K^{(-)}(\lambda)$ matrix. The strategy for $N > 3$ is
to substitute Eq.(\ref{Ksol}) into the reflection equation
(\ref{reflection}) and to search for further relations that are able
to determine the remaining ratios among next-neighbor amplitudes. By performing
this analysis up to $N=5$ we observed that some of the functional
equations can be fullfiled once we extend the ansatz (\ref{Ksol}) to
any ratio $\frac{K_{a+1}^{(-)}(\lambda)}{K_{a}^{(-)}(\lambda)}$,
\EQ
\frac{K_{a+1}^{(-)}(\lambda)}{K_{a}^{(-)}(\lambda)} =
\frac{\sinh[\beta_{a}-\lambda]}{\sinh[\beta_{a}+\lambda]},
~~a=1,\dots,N-1. \label{KsolA}
\EN

Now by substituting this proposal back to the reflection equation
and after systematic algebraic manipulations, we find that the
parameters $\beta_a$ have to satisfy the following recurrence
relation
\EQ
\beta_{a+1}-\beta_a=\frac{\IM \pi k}{N}, ~~a=1,\dots,N-1.
\EN

Putting together all the above results we conclude that the most
general solution for the amplitudes are,
\EQ K_a^{(-)}(\lambda) =
\prod_{b=1}^{a-1} \sinh[\xi_- + \IM \frac{\gamma}{2} - \frac{\IM \pi
k}{N}(\frac{1}{2}-b)-\lambda] \prod_{b=a}^{N-1} \sinh[\xi_- +
\IM \frac{\gamma}{2} - \frac{\IM \pi k}{N}(\frac{1}{2}-b)+\lambda]
\label{KweightM} \EN
where for later convenience we choose
$\beta_1=\xi_-+\IM \frac{\gamma}{2}+\frac{\IM \pi k}{2N}$ such that the
variable $\xi_-$ is a free continuous parameter. We also emphasize
that we have checked that the K-matrix (\ref{KweightM}) satisfy
the reflection equation (\ref{reflection}) until $N=7$.

Considering the isomorphism (\ref{isomorphism}) one can easily
derive that the form of the respective elements of
$K^{(+)}(\lambda)$ are given by
\EQ K_a^{(+)}(\lambda) =
\prod_{b=1}^{a-1} \sinh[\xi_+ + \IM \frac{\gamma}{2} - \frac{\IM \pi
k}{N}(\frac{3}{2}-b)+\lambda] \prod_{b=a}^{N-1} \sinh[\xi_+ +
\IM \frac{\gamma}{2} + \frac{\IM \pi k}{N}(\frac{1}{2}+b)-\lambda]
\label{KweightP} \EN
where $\xi_+$ is yet another free continuous
variable.

Having at hand the reflection matrices $K^{(\pm)}(\lambda)$ one can
construct an integrable model with open boundaries following the
double-row transfer matrix formulation proposed by Sklyanin
\cite{SK},
\EQ
t(\lambda) = Tr_{{\cal A}} \left[ K^{(+)}_{\cal
A}(\lambda) {\cal T}_{\cal A}(\lambda) K^{(-)}_{\cal A}(\lambda)
{\cal T}_{\cal A}(-\lambda)^{-1} \right] \EN
where ${\cal T}_{\cal
A}(\lambda)$ is the monodromy matrix of the corresponding closed
chain with $L$ sites,
\EQ {\cal T}_{\cal A}(\lambda) =
\bar{R}_{{\cal A} L}(\lambda) \bar{R}_{{\cal A} L-1}(\lambda) \dots
\bar{R}_{{\cal A} 1}(\lambda).
\EN

To obtain the respective Hamiltonian with open boundaries one needs
to expand the double-row transfer matrix $t(\lambda)$ in powers
$\lambda$. The first derivative of $t(\lambda)$ is proportional to
$Tr_{{\cal A}} \left[ K^{(+)}_{\cal A}(0) \right]$ which in our case
is null for arbitrary values of $\gamma$ and $N$. In this situation
we have to consider the expansion of $t(\lambda)$ up to the second
order in the spectral parameter $\lambda$. Considering that
$K^{(-)}_{\cal A}(\lambda)$ has been normalized such that
$K^{(-)}_{\cal A}(0) = I_{N}$, the expression for the Hamiltonian is
\cite{HAM},
\bear H &=& \sum_{j=1}^{L-1} H_{j,j+1}+ \frac{\rho(0)}{2} \left.
\frac{d}{d \lambda}K^{(+)}_1(\lambda) \right|_{\lambda=0} +
\frac{1}{\zeta} Tr_{{\cal A}} \left[ \left. \frac{d}{d
\lambda} K^{(+)}_{\cal A}(\lambda)\right|_{\lambda=0} H_{L,{\cal A}}
\right.
\nonumber \\
&+& \left. \frac{1}{2} K^{(+)}_{\cal A}(0) \left. \frac{d^2}{d
\lambda^2} R_{{\cal A} L}(\lambda) \right|_{\lambda=0} P_{L {\cal
A}} +\frac{1}{2 \rho(0)} K^{(+)}_{\cal A}(0) H_{L, {\cal A}}^2
\right], \ear
where $H_{j,j+1}$ is the standard bulk Hamiltonian
$H_{j,j+1} = P_{j,j+1} \left. \frac{d}{d \lambda} \bar{R}_{j
j+1}(\lambda) \right|_{\lambda=0}$ while $\zeta$ is a constant
proportional to the following identity matrix
\EQ
\zeta I_{N}
= Tr_{{\cal A}} \left[ \left. \frac{d}{d \lambda} K_{{\cal
A}}^{(+)}(\lambda) \right|_{\lambda=0} + \frac{2}{\rho(0)} K_{{\cal
A}}^{(+)}(0) H_{L,{\cal A}}\right]
\EN

Let us now present the explicit expressions of the open
Hamiltonians in the simplest cases $N = 2$ and $N = 3$. For $N=2$
we have the $XX$ chain in the presence of both bulk and boundary magnetic fields,
\bear
H = \sum_{i=1}^{L-1}
\frac{1}{2}(\sigma_{i}^x \sigma_{i+1}^x + \sigma_{i}^y \sigma_{i+1}^y) +
\frac{\cos[\gamma]}{2} \left(\sigma_{i}^z + \sigma_{i+1}^z  \right)
&+& \frac{\sin[\gamma]}{2} \cot \left[-\IM \xi_- + \frac{\gamma}{2} +
\frac{\pi}{4} \right] \sigma_{1}^z \nonumber \\
&+& \frac{\sin[\gamma]}{2} \cot \left[\IM \xi_+ + \frac{\gamma}{2} +
\frac{\pi}{4} \right] \sigma_{L}^z
 \ear
where $\sigma_{i}^{x}$, $\sigma_{i}^{y}$  and $\sigma_i^z$ are
Pauli matrices.

On the other hand for $N=3$ we find  that the corresponding open Hamiltonian up
to an additive constant is,
\bear H &=&
\frac{\sqrt{3}}{2}
\sum_{i=1}^{L-1}
\sin[\gamma \varepsilon_k+\frac{\pi}{6}] \left(S^x_i S^x_{i+1} + S^y_i
S^y_{i+1} \right) + \left(S^x_i S^x_{i+1} +
S^y_i S^y_{i+1} \right)^2
\nonumber \\
&+&
\left(
\sin[\gamma \varepsilon_k+\frac{\pi}{6}]
-2\sqrt{\frac{\sin[\gamma]
\sin[\gamma+\varepsilon_k \frac{\pi}{3}]}{3}} \right )
\left[ \left(S^x_i S^x_{i+1} + S^y_i S^y_{i+1} \right)
S^z_i S^z_{i+1} +h.c. \right ]
%
\nonumber \\
&+& \frac{1}{\sqrt{3}} \sin[-\gamma \varepsilon_k+\frac{\pi}{3}]
\left(S^x_i S^x_{i+1} + S^y_i S^y_{i+1} \right) \left( S^z_i +
S^z_{i+1} \right)
+ \frac{1}{2} S^z_i S^z_{i+1}
-\frac{1}{2} (S^z_i S^z_{i+1})^2
\nonumber \\
&+& \frac{1}{\sqrt{3}}\sin[2 \gamma + \frac{\pi k}{3}] (S^z_i + S^z_{i+1})
+ \frac{3}{2} (S^{z^2}_i + S^{z^2}_{i+1})
\nonumber \\
&+&
\frac{\sin[\gamma] \sin[\gamma + \frac{\pi k}{3}]}{2 \sin[-\IM \xi_-
+\frac{\gamma}{2} + \frac{\pi k}{6}] \sin[-\IM \xi_- +\frac{\gamma}{2}
+ \frac{ \pi k}{2}]} \left(
\sin[-2 \IM \xi_- + \gamma+\frac{2 \pi k}{3} ] S_{1}^z + \frac{\sqrt{3}}{2} \left(S_{1}^z\right)^2 \right)
\nonumber \\
&+&
\frac{\sin[\gamma] \sin[\gamma + \frac{\pi k}{3}]}{2 \sin[\IM \xi_+
+ \frac{\gamma}{2} + \frac{\pi k}{2}] \sin[\IM \xi_+ +
\frac{\gamma}{2} + \frac{ \pi k}{6}]} \left( \sin[2 \IM \xi_+ +\gamma
+ \frac{2 \pi k}{3}] S_{L}^z + \frac{\sqrt{3}}{2}
\left(S_{L}^z\right)^2 \right)
\ear
such that $S_i^{x}$, $S_i^{y}$ and $S_i^z$ denotes the standard $SU(2)$
spin-$1$ matrices.

In next section we shall consider the diagonalization of the double-row transfer
matrix associated to these models.

\section{Bethe ansatz analysis}

The purpose of this section is to present the eigenvalues
$\Lambda_n(\lambda)$ of the double-row transfer matrix,
\EQ
t(\lambda) \ket{\phi_n} = \Lambda_n(\lambda) \ket{\phi_n}.
\label{eigenvalueproblem}\EN

In the case of diagonal $K$-matrices the diagonalization problem
(\ref{eigenvalueproblem}) can be tackled within the algebraic Bethe
ansatz framework. It turns out that the standard ferromagnetic
highest vector $\ket{\phi_0}$ is an exact eigenvector of
$t(\lambda)$, \EQ \ket{\phi_0}=\prod_{i=1}^{L} \otimes \ket{0}_{i},
~~~~ \ket{0}_{i}=\left(\begin{array}{c} 1 \\ 0 \\ \vdots \\ 0
\end{array}\right)_{N} \EN
playing the role of suitable reference state to start the Bethe
ansatz analysis.

An algebraic procedure to generate other eigenvectors of
$t(\lambda)$ was first devised by Sklyanin for $N=2$ \cite{SK}. The
central object in this approach is the matrix elements of the
double-row monodromy matrix, \EQ \bar{\cal T}_{\cal A}(\lambda) =
{\cal T}_{\cal A}(\lambda) K_{\cal A}^{(-)}(\lambda) {\cal T}_{\cal
A}(\lambda)^{-1} \EN which also satisfies the reflection equation
(\ref{reflection}), \bear \bar{R}_{12}(\lambda-\mu) \bar{\cal
T}_{\cal A}^1(\lambda) \bar{R}_{21}(\lambda+\mu) \bar{\cal T}_{\cal
A}^2(\mu) = \bar{\cal T}_{\cal A}^2(\mu) \bar{R}_{12}(\lambda+\mu)
\bar{\cal T}_{\cal A}^1(\lambda) \bar{R}_{21}(\lambda-\mu)
\label{quadraticalgebra} \ear

The other eigenstates of $t(\lambda)$ were then constructed by
exploring a set of commutation relations derived from the quadratic
algebra (\ref{quadraticalgebra}). This algebraic analysis has been
extended to tackle three-state vertex models \cite{FAN,LIE} as well
as the isotropic Heisenberg chain with arbitrary $N$
up to the two-particle eigenstates
\cite{CGM}. In what follows we shall adapt the
results of the latter work to include vertex models
that are not invariant by charge
symmetry. We shall not repeat the technical details discussed in
\cite{CGM} but only present the main relevant steps necessary to
construct the eigenvalues $\Lambda_n(\lambda)$. In order to describe
that we shall first represent the double-monodromy matrix as 
\bear
\bar{\cal T}_{\cal A}(\lambda)= \left(\begin{array}{cccc}
\bar{\cal T}_{1,1}(\lambda) & \bar{\cal T}_{1,2}(\lambda) & \dots & \bar{\cal T}_{1,N}(\lambda) \\
\bar{\cal T}_{2,1}(\lambda) & \bar{\cal T}_{2,2}(\lambda) & \dots & \bar{\cal T}_{2,N}(\lambda) \\
\vdots & \vdots & \ddots & \vdots \\
\bar{\cal T}_{N,1}(\lambda) & \bar{\cal T}_{N,2}(\lambda) & \dots & \bar{\cal T}_{N,N}(\lambda) \\
\end{array}\right) \ear

Taking into account this representation, the diagonalization of the
double-row transfer matrix $t(\lambda)$ becomes equivalent to the
problem \EQ \sum_{i=1}^{N} K_i^{(+)}(\lambda) \bar{\cal
T}_{i,i}(\lambda) \ket{\phi_n} = \Lambda_n(\lambda) \ket{\phi_n}
\label{doubleeigeinvalueproblem}\EN

The first step in the algebraic framework is to reformulate
(\ref{doubleeigeinvalueproblem})
in
terms of a suitable combination of diagonal fields $\bar{\cal
T}_{i,i}(\lambda)$. It turns out that for arbitrary $N$ such
linear combination is given by,
\EQ \bar{\cal T}_{i,i}(\lambda)=
\sum_{j=1}^{i} \frac{\mid M_{j,i}^{(+)} (2 \lambda) \mid}{{\mid
M_{j,j}^{(+)} (2 \lambda) \mid}} \bar{\cal T}_{j,j}^{'}(\lambda)
\label{translation} \EN
where the $j \times j$ matrix
$M_{j,i}^{(+)}(\lambda)$ is build up from the $R$-matrix elements
$\bar{R}_{1,2}(\lambda)$ by the expression,
\bear
M_{j,i}^{(+)}(\lambda)= \left[\begin{array}{ccccc}
\bar{R}_{1,1}^{1,1}(\lambda) & \bar{R}_{2,1}^{1,2}(\lambda) & \dots & \bar{R}_{j-1,1}^{1,j-1}(\lambda) & \bar{R}_{i,1}^{1,i}(\lambda) \\
\bar{R}_{1,2}^{2,1}(\lambda) & \bar{R}_{2,2}^{2,2}(\lambda) & \dots & \bar{R}_{j-1,2}^{2,j-1}(\lambda) & \bar{R}_{i,2}^{2,i}(\lambda) \\
\vdots & \vdots & \ddots & \vdots & \vdots \\
\bar{R}_{1,j}^{j,1}(\lambda) & \bar{R}_{2,j}^{j,2}(\lambda) & \dots & \bar{R}_{j-1,j}^{j,j-1}(\lambda) & \bar{R}_{i,j}^{j,i}(\lambda) \\
\end{array}\right]_{j \times j} \label{AUXmatrix1} \ear

Now by using Eqs(\ref{doubleeigeinvalueproblem},\ref{translation})
the eigenvalue problem can be rewritten as, 
\EQ \sum_{i=1}^{N}
w_i^{(+)}(\lambda) \bar{\cal T}_{i,i}^{'}(\lambda) \ket{\phi_n} =
\Lambda_n(\lambda) \ket{\phi_n} \label{finaleigeinvalueproblem} 
\EN
where functions $w_i^{(+)}(\lambda)$ are defined by \EQ
w_i^{(+)}(\lambda) = \sum_{j=i}^{N} \frac{\mid M_{i,j}^{(+)} (2
\lambda) \mid}{{\mid M_{i,i}^{(+)} (2 \lambda) \mid}}
K_j^{(+)}(\lambda). \EN

The basic property of the new diagonal fields $\bar{\cal
T}_{i,i}^{'}(\lambda)$ is that their action on the reference
$\ket{\phi_0}$ are always proportional to functions having a single
power in $L$. More precisely we find that, 
\EQ \bar{\cal
T}_{i,i}^{'}(\lambda) \ket{\phi_0}  = w_i^{(-)}(\lambda)
\frac{\left[\bar{R}_{i,1}^{i,1}(\lambda)\right]^{2L}}{\left[
\rho(\lambda) \rho(-\lambda) \right]^L} \ket{\phi_0} 
\label{change}
\EN

The functions $w_i^{(-)}(\lambda)$ depend now on the amplitudes
$K_i^{(-)}(\lambda)$ by the expression, \EQ w_i^{(-)}(\lambda) =
K_i^{(-)}(\lambda) - \sum_{j=1}^{i-1} \frac{\mid M_{i-1,j}^{(-)} (2
\lambda) \mid}{{\mid M_{i-1,i-1}^{(+)} (2 \lambda) \mid}}
K_j^{(-)}(\lambda) \EN where the second auxiliary $j \times j$ matrix
$M_{j,i}^{(-)}(\lambda)$ is also build up from the $R$-matrix
elements $\bar{R}_{1,2}(\lambda)$, 
\bear
M_{j,i}^{(-)}(\lambda)= \left[\begin{array}{ccccccc}
\bar{R}_{1,1}^{1,1}(\lambda) & \dots & \bar{R}_{i-1,1}^{1,i-1}(\lambda) & \bar{R}_{j+1,1}^{1,j+1}(\lambda) & \bar{R}_{i+1,1}^{1,i+1}(\lambda) & \dots & \bar{R}_{j,1}^{1,j}(\lambda) \\
\bar{R}_{1,2}^{2,1}(\lambda) & \dots & \bar{R}_{i-1,2}^{2,i-1}(\lambda) & \bar{R}_{j+1,2}^{2,j+1}(\lambda) & \bar{R}_{i+1,2}^{2,i+1}(\lambda) & \dots & \bar{R}_{j,2}^{2,j}(\lambda) \\
\vdots & \ddots & \vdots & \vdots & \vdots & \ddots & \vdots \\
\bar{R}_{1,j}^{j,1}(\lambda) & \dots & \bar{R}_{i-1,j}^{j,i-1}(\lambda) & \bar{R}_{j+1,j}^{j,j+1}(\lambda) & \bar{R}_{i+1,j}^{j,i+1}(\lambda) & \dots & \bar{R}_{j,j}^{j,j}(\lambda) \\
\end{array}\right]_{j \times j} \label{AUXmatrix2} \ear

To this point we have gathered the basic informations to provide a
closed expression for the eigenvalue $\Lambda_0(\lambda)$ associated
to the reference state $\ket{\phi_0}$. In fact, by substituting
Eq.(\ref{change}) in Eq.(\ref{finaleigeinvalueproblem}) one
finds, \EQ \Lambda_0(\lambda) = \sum_{i=1}^{N} w_i^{(+)}(\lambda)
w_i^{(-)}(\lambda)
\frac{\left[\bar{R}_{i,1}^{i,1}(\lambda)\right]^{2L}}{\left[
\rho(\lambda) \rho(-\lambda) \right]^L}. \EN

To obtain the other eigenvalues of $t(\lambda)$ we have to look for
eigenstates generated by the action of the creation field $\bar{\cal
T}_{i,j}(\lambda)$ for $i<j$ on the reference state $\ket{\phi_0}$.
This analysis has been already carried out in \cite{CGM} up to the
two-particle states. Recall that in integrable theories the results
for the two-particle state are enough to propose an educated
expression for the eigenvalues $\Lambda_n(\lambda)$. By adapting the
conclusion of \cite{CGM} to our vertex model we find that the
expression for the multi-particle eigenvalues are
\EQ
\Lambda_n(\lambda) = \sum_{i=1}^{N}
\frac{\left[\bar{R}_{i,1}^{i,1}(\lambda)\right]^{2L}}{\left[
\rho(\lambda) \rho(-\lambda) \right]^L} w_i^{(+)}(\lambda)
w_i^{(-)}(\lambda)  \prod_{j=1}^{n} Q_i(\lambda,\lambda_j) \EN
where $\lambda_1,\cdots,\lambda_n$ are the variables parameterizing
the $n$-particle state.

The functions $Q_i(\lambda,\lambda_j)$ are determined in terms of the
Boltzmann weights by the following expression,
\EQ
Q_i(\lambda,\mu) = \begin{cases} \displaystyle
\frac{\bar{R}_{1,1}^{1,1}(\mu-\lambda)}{\bar{R}_{1,2}^{1,2}(\mu-\lambda)}
\frac{\bar{R}_{1,2}^{1,2}(\lambda+\mu)}{\bar{R}_{1,1}^{1,1}(\lambda+\mu)}
, ~~~~\mbox{for} ~~ i=1 \cr \displaystyle \frac{\left|
\begin{array}{cc} \bar{R}_{1,i+1}^{1,i+1}(\lambda-\mu) &
\bar{R}_{1,i+1}^{2,i}(\lambda-\mu) \\
\bar{R}_{2,i}^{1,i+1}(\lambda-\mu) &
\bar{R}_{2,i}^{2,i}(\lambda-\mu)
\end{array} \right|}{\bar{R}_{1,i+1}^{1,i+1}(\lambda-\mu) \bar{R}_{1,i}^{1,i}(\lambda-\mu)}
\frac{\left|
\begin{array}{cc} \bar{R}_{i-1,1}^{i-1,1}(\lambda+\mu) &
\bar{R}_{i-1,2}^{i,1}(\lambda+\mu) \\
\bar{R}_{i,1}^{i-1,2}(\lambda+\mu) &
\bar{R}_{i,2}^{i,2}(\lambda+\mu)
\end{array} \right|}{\bar{R}_{i-1,1}^{i-1,1}(\lambda+\mu) \bar{R}_{1,i}^{1,i}(\lambda+\mu)}
, ~~~~ \mbox{for} ~~ 2 \le  i \le N-1 \cr \displaystyle \frac{
\bar{R}_{2,N}^{2,N}(\lambda-\mu)}{\bar{R}_{N,1}^{N,1}(\lambda-\mu)}
\frac{\left|
\begin{array}{cc} \bar{R}_{N-1,1}^{N-1,1}(\lambda+\mu) &
\bar{R}_{N,1}^{N-1,2}(\lambda+\mu) \\
\bar{R}_{N-1,2}^{N,1}(\lambda+\mu) &
\bar{R}_{N,2}^{N,2}(\lambda+\mu)
\end{array} \right|}{\bar{R}_{N-1,1}^{N-1,1}(\lambda+\mu) \bar{R}_{1,N}^{1,N}(\lambda+\mu)}
~~~~\mbox{for} ~~ i = N
\end{cases}
\EN
while the rapidities $\lambda_j$ satisfy the following Bethe
ansatz equations,
\EQ \left[
\frac{\bar{R}_{1,1}^{1,1}(\lambda_j)}{\bar{R}_{2,1}^{2,1}(\lambda_j)}
\right]^{2L} \frac{w_1^{(+)}(\lambda_j)}{w_2^{(+)}(\lambda_j)}
\frac{w_1^{(-)}(\lambda_j)}{w_2^{(-)}(\lambda_j)} =
\frac{\bar{R}_{1,1}^{1,1}(2 \lambda_j) \bar{R}_{2,2}^{2,2}(2
\lambda_j) - \bar{R}_{2,1}^{1,2}(2 \lambda_j) \bar{R}_{1,2}^{2,1}(2
\lambda_j)}{ \left[ \bar{R}_{2,1}^{2,1}(2 \lambda_j) \right]^2 }
\prod_{\stackrel{i=1}{i \ne j}}^{n}
\frac{Q_2(\lambda_j,\lambda_i)}{Q_1(\lambda_j,\lambda_i)}
 \EN
for $j=1,\cdots, n$.

We now have at hand the basic ingredients to exhibit explicit
expressions for the eigenvalues $\Lambda_n(\lambda)$. The first step
is to simplify the expressions for $w_1^{(\pm)}(\lambda),
\cdots,
w_N^{(\pm)}(\lambda)
$  by using
the $K$-matrices amplitudes (\ref{KweightM},\ref{KweightP}) as well
as the $R$-matrix elements provided in section 2 and Appendix B.
These simplifications require a considerable amount of algebraic
work since we have to sum a number of distinct terms for each
possible branch $k$. 
Fortunately, an analysis up to $N=5$ is enough to exhibit the uniform
dependence
of these functions on $N$.
We find that the final results factorize in
terms of three types of products of trigonometric functions, namely
\bear
w_a^{(-)}(\lambda) &=& \prod_{j=1}^{a-1} \frac{\sinh[2 \lambda
+ \frac{\IM \pi k}{N} (j-1)]}{\sinh[2 \lambda + \IM \gamma + \frac{\IM \pi
k}{N} (a+j-3)]} \prod_{j=1}^{a-1} \sinh[\xi_- - \lambda -\IM
\frac{\gamma}{2} - \frac{\IM \pi k}{N} (j-\frac{3}{2})] \nonumber
\\
&\times & \prod_{j=a}^{N-1} \sinh[\xi_- + \lambda + \IM \frac{\gamma}{2}
- \frac{\IM \pi k}{N} (\frac{1}{2}-j)] 
\ear
and
\bear
w_a^{(+)}(\lambda) &=& \prod_{j=a}^{N-1} \frac{\sinh[2 \lambda +
\frac{\IM \pi k}{N} (j-1)]}{\sinh[2 \lambda + \IM \gamma + \frac{\IM \pi
k}{N} (a+j-2)]} \prod_{j=1}^{a-1} \sinh[\xi_+ + \lambda +
\IM \frac{\gamma}{2} - \frac{\IM \pi k}{N} (\frac{3}{2}-j)] \nonumber
\\
&\times & \prod_{j=a}^{N-1} \sinh[\xi_+ - \lambda -\IM \frac{\gamma}{2}
+ \frac{\IM \pi k}{N} (N+\frac{1}{2}-j)]
\ear

The next step is to carry out similar algebraic simplifications
for functions $Q_i(\lambda,\lambda_j)$.
We notice that in order to make these polynomials
as symmetrical as possible it is convenient to perform the
shift $\lambda_i \rightarrow
\bar{\lambda}_i-\IM \frac{\gamma}{2}$. Considering this change
of variables and
after some manipulations we
find that the double-row transfer matrix eigenvalue are,
\bear
\Lambda_n(\lambda) &=& \sum_{a=1}^{N}
\left[\frac{\rho(-\lambda)}{\rho(\lambda)} \right]^L \left[
\prod_{j=1}^{a-1} \frac{\sinh[\lambda + \frac{\IM \pi k}{N}
(j-1)]}{\sinh[\lambda + \IM \gamma + \frac{\IM \pi k}{N} (j-1)]}
\right]^{2L} w_a^{(+)}(\lambda) w_a^{(-)}(\lambda) \nonumber \\
&\times & \prod_{i=1}^{n} \frac{\sinh[\lambda - \bar{\lambda}_i -
\IM \frac{\gamma}{2}] \sinh[\lambda - \bar{\lambda}_i + \IM \frac{\gamma}{2} -
\frac{\IM \pi k}{N}]}{\sinh[\lambda - \bar{\lambda}_i + \IM \frac{\gamma}{2} -
(2-a) \frac{\IM \pi k}{N}] \sinh[\lambda - \bar{\lambda}_i +
\IM \frac{\gamma}{2} - (1-a) \frac{\IM \pi k}{N}]}
\nonumber \\
&\times & \prod_{i=1}^{n} \frac{\sinh[\lambda + \bar{\lambda}_i -
\IM \frac{\gamma}{2}] \sinh[\lambda + \bar{\lambda}_i + \IM \frac{\gamma}{2} -
\frac{\IM \pi k}{N}]}{\sinh[\lambda + \bar{\lambda}_i + \IM \frac{\gamma}{2} -
(2-a) \frac{\IM \pi k}{N}] \sinh[\lambda + \bar{\lambda}_i +
\IM \frac{\gamma}{2} - (1-a) \frac{\IM \pi k}{N}]}
\ear

The corresponding Bethe ansatz equations for the shifted variables
$\bar{\lambda}_i$ become,

\bear \left( \frac{\sinh[\bar{\lambda}_i + \IM
\frac{\gamma}{2}]}{\sinh[\bar{\lambda}_i - \IM \frac{\gamma}{2}]} \right)^{2L}
\frac{\sinh[\xi_- + \bar{\lambda}_i + \frac{\IM \pi k }{2 N}]} {\sinh[\xi_-
- \bar{\lambda}_i + \frac{\IM \pi k }{2 N}]} \frac{\sinh[\xi_+ - \bar{\lambda}_i -
\frac{\IM \pi k}{2N}]} {\sinh[\xi_+ + \bar{\lambda}_i - \frac{\IM \pi
k}{2N}]}
 \nonumber \\
=  \prod_{\stackrel{j=1}{j \ne i}}^n \frac{\sinh[\bar{\lambda}_i -
\bar{\lambda}_j - \frac{\IM \pi k}{N}]} {\sinh[\bar{\lambda}_i - \bar{\lambda}_j +
\frac{\IM \pi k}{N}]} \frac{\sinh[\bar{\lambda}_i + \bar{\lambda}_j - \frac{\IM \pi
k}{N}]} {\sinh[\bar{\lambda}_i + \bar{\lambda}_j
 + \frac{\IM \pi k}{N}]}
\label{bethe}
\ear

We conclude with the following remark. Note that the roots of unity
branches on the left-hand side of Eq.(\ref{bethe}) can be absorbed
by performing the change of variables $\xi_- \rightarrow \xi_- -
\frac{\IM \pi k }{2 N}$ and $\xi_+ \rightarrow \xi_+ + \frac{\IM \pi
k}{2N}$. After this transformation, the dependence of the Bethe
ansatz equations  on the roots of unity remains restricted to the
two-body scattering amplitudes.

\section{Conclusion}

The purpose of this paper was to solve the integrable vertex models based
on the $U_q[SU(2)]$ algebra at roots of unity with open boundary 
conditions. We have solved the reflection equation and found
one family of diagonal $K$-matrices having a free-parameter.
For such diagonal boundary conditions we have been able to present
the eigenvalues of the double-row transfer matrix and the corresponding
Bethe ansatz equations. The next natural step would be to consider these
vertex models with non-diagonal boundaries. In particular, to investigate
if the functional relation approach developed for the open high spin
$XXZ$ quantum chain \cite{NEPX,WY,DOI,MUR}  can be applied to
such roots of unity vertex models with non-diagonal $K$-matrices.

\addcontentsline{toc}{section}{Appendix A}
\section*{\bf Appendix A: The projectors $P_j(\gamma,k)$} \setcounter{equation}{0}
\renewcommand{\theequation}{A.\arabic{equation}}

For completeness we shall here present the explicit expressions for
the projectors $P_j(\gamma,k)$ given first by us in \cite{MC}. In
the Weyl basis these projectors can be written as decomposition of
the braid $S(\gamma,k)$,
\EQ P_j(\gamma,k) = \prod_{\stackrel{l=1}{l
\ne j}}^{N} \frac{S(\gamma,k) - \xi_l}{\xi_j-\xi_l} 
\label{proj1}
\EN where \EQ
\xi_j=(-1)^{j+1} e^{\IM [\frac{\pi k}{N}(j-2)(j-1)+2 \gamma (j-1)]}.
\EN

The corresponding matrix representation of the braid $S(\gamma,k)$
is 
\EQ
S(\gamma,k)= \sum_{\stackrel{a,b,c,d=1}{a \ge d; c \ge b }}^{N}
S_{c,d}^{a,b}(\gamma,k)
e_{b,d}
\otimes
e_{a,c}.
\EN
where
the amplitudes $S^{c,d}_{a,b}$ are given by, 
\bear
S_{c,d}^{a,b}(\gamma,k) & =& \frac{\exp[\frac{2 \pi \IM k}{N}(b-1)(d-1) +\IM \gamma(b+d-2)]}
{H(\frac{2 \pi k}{N},a-d) }
\sqrt{
\frac{ H(\frac{2 \pi k}{N},a-1) H(\frac{2 \pi k}{N},c-1)}
{H(\frac{2 \pi k}{N},d-1) H(\frac{2 \pi k}{N},b-1)}}
\nonumber \\ 
& \times &
\sqrt{
\frac{H(2\gamma,a-1) H(2 \gamma,c-1)}
{H(2 \gamma,d-1) H(2 \gamma,b-1)}}
\delta_{a+b,c+d},
\label{proj2}
\ear
such
that the auxiliary function $H(\lambda,n) =
\displaystyle{\prod_{l=0}^{n-1}} (1-e^{\IM [\lambda + \frac{2 \pi
k}{N} l]})$.

Altogether the above expressions provide us the explicit matrix
expressions for the projectors for arbitrary and $k$ coprime with
$N$.

\addcontentsline{toc}{section}{Appendix B}
\section*{\bf Appendix B: Boltzmann weights for $N=4$ and $N=5$}
\setcounter{equation}{0}
\renewcommand{\theequation}{B.\arabic{equation}}

In what follows we present the weights
$\bar{R}_{a,b}^{c,d}(\lambda)$ of the $R$-matrix
$\bar{R}_{12}(\lambda)$ (\ref{Rmatrix}) for $N=4$ and $N=5$. We
recall here that the only non-null amplitudes are those that satisfy
the ice rule $a+b=c+d$.

The non-trivial forty-four Boltzmann weigths for $N=4$ are,
\bear
R_{1,1}^{1,1}(\lambda) &=& \sinh[\IM \gamma + \lambda] \sinh[\IM \gamma +
\frac{\IM  \pi k}{4} + \lambda] \sinh[\IM \gamma + \frac{\IM \pi k}{2} +
\lambda]
\\
R_{1,2}^{1,2}(\lambda) &=& R_{2,1}^{2,1}(\lambda) = \sinh[\lambda]
\sinh[\IM \gamma + \frac{\IM \pi k}{4} + \lambda] \sinh[\IM \gamma + \frac{\IM
\pi k}{2} + \lambda]
\\
R_{1,2}^{2,1}(\lambda) &=& R_{2,1}^{1,2}(\lambda) = \sinh[\IM \gamma]
\sinh[\IM \gamma + \frac{\IM \pi k}{4} + \lambda] \sinh[\IM \gamma + \frac{\IM
\pi k}{2} + \lambda]
\\
R_{1,3}^{1,3}(\lambda) &=& R_{3,1}^{3,1}(\lambda) = \sinh[\lambda]
\sinh[\lambda + \frac{\IM \pi k}{4}] \sinh[\IM \gamma + \frac{\IM \pi k}{2}
+ \lambda]
\\
R_{1,3}^{2,2}(\lambda) &=& R_{2,2}^{1,3}(\lambda) =
R_{2,2}^{3,1}(\lambda) = R_{3,1}^{2,2}(\lambda) = -{\frac{2}{\varepsilon_k}}^{\frac{1}{4}}
\sqrt{\sinh[\IM \gamma] \sinh[\IM \gamma +
\frac{\IM \pi k}{4}]} \nonumber \\
&\times& \sinh[\IM \gamma] \sinh[\IM \gamma + \frac{\IM \pi k}{2} + \lambda]
%
%
\\
R_{1,3}^{3,1}(\lambda) &=& R_{3,1}^{1,3}(\lambda) = \sinh[\IM \gamma]
\sinh[\IM \gamma + \frac{\IM \pi k}{4}] \sinh[\IM \gamma + \frac{\IM \pi k}{2} +
\lambda]
\\
R_{1,4}^{1,4}(\lambda) &=& R_{4,1}^{4,1}(\lambda) = \sinh[\lambda +
\frac{\IM \pi k}{4}] \sinh[\lambda + \frac{\IM \pi k}{2}] \sinh[\lambda]
\ear
\bear
%
R_{1,4}^{2,3}(\lambda) &=& R_{2,3}^{1,4}(\lambda) =
R_{3,2}^{4,1}(\lambda) = R_{4,1}^{3,2}(\lambda) =
-{\varepsilon_k}^{\frac{1}{2}}
\sqrt{\sinh[\IM \gamma] \sinh[\IM \gamma + \frac{\IM \pi k}{2}]}
\sinh[\lambda] \sinh[\lambda + \frac{\IM \pi k}{4}]
\nonumber \\
\ear
\bear
R_{1,4}^{3,2}(\lambda) &=& R_{3,2}^{1,4}(\lambda) =
R_{2,3}^{4,1}(\lambda) = R_{4,1}^{2,3}(\lambda) =
-{\varepsilon_k}^{\frac{1}{2}}
\sqrt{\sinh[\IM \gamma] \sinh[\IM \gamma + \frac{\IM \pi k}{2}]}
\sinh[\lambda] \sinh[\IM \gamma + \frac{\IM \pi k}{4}]
\nonumber \\
\ear
\bear
R_{1,4}^{4,1}(\lambda) &=& R_{4,1}^{1,4}(\lambda) = \sinh[\IM \gamma]
\sinh[\IM \gamma + \frac{\IM \pi k}{4}] \sinh[\IM \gamma + \frac{\IM \pi k}{2}]
\\
R_{2,2}^{2,2}(\lambda) &=& \left[ \sinh[\IM \gamma] \sinh[\IM \gamma +
\frac{\IM  \pi k}{4}] + \sinh[\lambda - \frac{\IM \pi k}{4}]
\sinh[\lambda] \right] \sinh[\IM \gamma + \frac{\IM \pi k}{2} + \lambda]
\ear
\bear
%
%
R_{2,3}^{2,3}(\lambda) &=& R_{3,2}^{3,2}(\lambda) = \left[ \sinh[2\IM
\gamma] \sinh[\frac{3 \IM \pi k}{4}] + \sinh[\lambda - \frac{\IM \pi
k}{4}] \sinh[\lambda] \right] \sinh[\lambda]
\\
R_{2,3}^{3,2}(\lambda) &=& R_{3,2}^{2,3}(\lambda) = \left[
\sinh[\IM \gamma] \sinh[\IM \gamma + \frac{\IM \pi k}{2}]
+ \sqrt{2\varepsilon_k} \sinh[\lambda - \frac{\IM \pi k}{4}] \sinh[\lambda] \right]
\nonumber \\
&\times & \sinh[\IM \gamma + \frac{\IM \pi k}{4}]
\\
R_{2,4}^{2,4}(\lambda) &=& R_{4,2}^{4,2}(\lambda) = \sinh[\lambda]
\sinh[\lambda + \frac{\IM \pi k}{4}] \sinh[\IM \gamma - \lambda]
\\
R_{2,4}^{3,3}(\lambda) &=& R_{3,3}^{2,4}(\lambda) =
R_{3,3}^{4,2}(\lambda) = R_{4,2}^{3,3}(\lambda) = ({2\varepsilon_k})^{\frac{1}{4}}
\sinh[\lambda] \sqrt{\sinh[\IM \gamma +
\frac{\IM \pi k}{4}] \sinh[\IM \gamma + \frac{\IM \pi k}{2}]} \nonumber \\
&\times& \sinh[\IM \gamma - \lambda] \sinh[\lambda]
\\
R_{2,4}^{4,2}(\lambda) &=& R_{4,2}^{2,4}(\lambda) = \sinh[\IM \gamma +
\frac{\IM \pi k}{4}] \sinh[\IM \gamma + \frac{\IM \pi k}{2}] \sinh[\IM \gamma -
\lambda]
\\
R_{3,3}^{3,3}(\lambda) &=& \sinh[\IM \gamma - \lambda] \left[
\sinh[\IM \gamma + \frac{\IM \pi k}{4}] \sinh[\IM \gamma + \frac{\IM \pi k}{2}]
+ \sinh[\lambda - \frac{\IM \pi k}{4}] \sinh[\lambda] \right]
\\
R_{3,4}^{3,4}(\lambda) &=& R_{4,3}^{4,3}(\lambda) = \sinh[\IM \gamma -
\lambda] \sinh[\IM \gamma + \frac{\IM \pi k}{4} - \lambda] \sinh[\lambda]
\\
R_{3,4}^{4,3}(\lambda) &=& R_{4,3}^{3,4}(\lambda) = \sinh[\IM \gamma -
\lambda] \sinh[\IM \gamma + \frac{\IM \pi k}{4} - \lambda] \sinh[\IM \gamma +
\frac{\IM \pi k}{2}]
\\
R_{4,4}^{4,4}(\lambda) &=& \sinh[\IM \gamma - \lambda] \sinh[\IM
\gamma + \frac{\IM \pi k}{4} - \lambda] \sinh[\IM \gamma + \frac{\IM
\pi k}{2} - \lambda] 
\ear where here we have $k=1$ or $k=3$. 

For $N=5$ the eighty-five Boltzmann weights are,

\bear
R_{1,1}^{1,1}(\lambda) &=& \sinh[\IM \gamma + \lambda] \sinh[\IM \gamma +
\frac{\IM \pi k}{5} + \lambda] \sinh[\IM \gamma + \frac{2 \IM \pi k}{5} +
\lambda] \sinh[\IM \gamma + \frac{3 \IM \pi k}{5} + \lambda]
\ear
\bear
R_{1,2}^{1,2}(\lambda) &=& R_{2,1}^{2,1}(\lambda) = \sinh[\lambda]
\sinh[\IM \gamma + \frac{\IM \pi k}{5} + \lambda] \sinh[\IM \gamma + \frac{2 I
\pi k}{5} + \lambda] \sinh[\IM \gamma + \frac{3 \IM \pi k}{5} + \lambda]
\nonumber \\
\ear
\bear
R_{1,2}^{2,1}(\lambda) &=& R_{2,1}^{1,2}(\lambda) = \sinh[\IM \gamma]
\sinh[\IM \gamma + \frac{\IM \pi k}{5} + \lambda] \sinh[\IM \gamma + \frac{2 I
\pi k}{5} + \lambda] \sinh[\IM \gamma + \frac{3 \IM \pi k}{5} + \lambda]
\nonumber \\
\ear
\bear
R_{1,3}^{1,3}(\lambda) &=& R_{3,1}^{3,1}(\lambda) = \sinh[\lambda]
\sinh[\lambda + \frac{\IM \pi k}{5}] \sinh[\IM \gamma + \frac{2 \IM \pi
k}{5} + \lambda] \sinh[\IM \gamma + \frac{3 \IM \pi k}{5} + \lambda]
\\
R_{1,3}^{2,2}(\lambda) &=& R_{2,2}^{1,3}(\lambda) =
R_{2,2}^{3,1}(\lambda) = R_{3,1}^{2,2}(\lambda) = - \sqrt{2
\cosh[\frac{\IM \pi k}{5}] \sinh[\IM \gamma]
\sinh[\IM \gamma + \frac{\IM \pi k}{5}]} \nonumber \\
&\times& \sinh[\lambda] \sinh[\IM \gamma + \frac{2 \IM \pi k}{5} +
\lambda] \sinh[\IM \gamma + \frac{3 \IM \pi k}{5} + \lambda]
\\
R_{1,3}^{3,1}(\lambda) &=& R_{3,1}^{1,3}(\lambda) = \sinh[\IM \gamma]
\sinh[\IM \gamma + \frac{\IM \pi k}{5}] \sinh[\IM \gamma + \frac{2 \IM \pi k}{5}
+ \lambda] \sinh[\IM \gamma + \frac{3 \IM \pi k}{5} + \lambda]
\\
R_{1,4}^{1,4}(\lambda) &=& R_{4,1}^{4,1}(\lambda) = \sinh[\lambda]
\sinh[\lambda + \frac{\IM \pi k}{5}] \sinh[\lambda + \frac{2 \IM \pi
k}{5}] \sinh[\IM \gamma + \frac{3 \IM \pi k}{5} + \lambda]
\\
R_{1,4}^{2,3}(\lambda) &=& R_{2,3}^{1,4}(\lambda) =
R_{3,2}^{4,1}(\lambda) = R_{4,1}^{3,2}(\lambda) = - \sinh[\lambda] \sinh[\lambda + \frac{\IM \pi k}{5}]
\sinh[\IM \gamma + \frac{3 \IM \pi k}{5} + \lambda]
\nonumber \\
& \times & \sqrt{2 \varepsilon_k \sinh[\IM \gamma] \sinh[\IM \gamma
+ \frac{2 \IM \pi k}{5}]\cosh[\frac{\IM \pi k}{5}]}
\\
R_{1,4}^{3,2}(\lambda) &=& R_{3,2}^{1,4}(\lambda) =
R_{2,3}^{4,1}(\lambda) = R_{4,1}^{2,3}(\lambda) = - \sinh[\lambda] \sinh[\IM \gamma + \frac{\IM \pi k}{5}]
\sinh[\IM \gamma + \frac{3 \IM \pi k}{5} + \lambda]
\nonumber \\
& \times & \sqrt{2 \varepsilon_k \sinh[\IM \gamma] \sinh[\IM \gamma
+ \frac{2 \IM \pi k}{5}] \cosh[\frac{\IM \pi k}{5}]}
\\
R_{1,4}^{4,1}(\lambda) &=& R_{4,1}^{1,4}(\lambda) = \sinh[\IM \gamma]
\sinh[\IM \gamma + \frac{\IM \pi k}{5}] \sinh[\IM \gamma + \frac{2 \IM \pi
k}{5}] \sinh[\IM \gamma + \frac{3 \IM \pi k}{5} + \lambda]
\\
R_{1,5}^{1,5}(\lambda) &=& R_{5,1}^{5,1}(\lambda) = \sinh[\lambda]
\sinh[\lambda+\frac{\IM \pi k}{5}] \sinh[\lambda+\frac{2 \IM \pi k}{5}]
\sinh[\lambda+\frac{3 \IM \pi k}{5}]
\\
R_{1,5}^{2,4}(\lambda) &=& R_{2,4}^{1,5}(\lambda) =
R_{5,1}^{4,2}(\lambda) = R_{4,2}^{5,1}(\lambda)
=-\sqrt{\varepsilon_k \sinh[\IM \gamma] \sinh[\IM \gamma+3 \frac{\IM
\pi k}{5}]} \sinh[\lambda]
\\
& \times & \sinh[\lambda+\frac{\IM \pi k}{5}] \sinh[\lambda+2
\frac{\IM \pi k}{5}]
\ear
\bear
R_{1,5}^{3,3}(\lambda) &=& R_{3,3}^{1,5}(\lambda) =
R_{5,1}^{3,3}(\lambda) = R_{3,3}^{5,1}(\lambda) =- 2 \sinh[\lambda]
\sinh[\lambda+\frac{\IM \pi k}{5}]
\nonumber \\
& \times & \sqrt{\varepsilon_k \sinh[\IM \gamma] \sinh[\IM
\gamma+\frac{\IM \pi k}{5}] \sinh[\IM \gamma+2 \frac{\IM \pi k}{5}]
\sinh[\IM \gamma+3 \frac{\IM \pi k}{5}] \cosh[\frac{\IM \pi k}{5}]
\cosh[\frac{2 \IM \pi k}{5}] }
\nonumber \\
\ear
\bear
R_{1,5}^{4,2}(\lambda) &=& R_{4,2}^{1,5}(\lambda) =
R_{5,1}^{2,4}(\lambda) = R_{2,4}^{5,1}(\lambda)
=-\sqrt{\varepsilon_k \sinh[\IM \gamma] \sinh[\IM \gamma+3 \frac{\IM
\pi k}{5}]} \sinh[\lambda]
\nonumber \\
& \times & \sinh[\IM \gamma+\frac{\IM \pi k}{5}] \sinh[\IM \gamma+2
\frac{\IM \pi k}{5}]
\\
R_{1,5}^{5,1}(\lambda) &=& R_{5,1}^{1,5}(\lambda) = \sinh[\IM \gamma]
\sinh[\IM \gamma+\frac{\IM \pi k}{5}] \sinh[\IM \gamma+2 \frac{\IM \pi k}{5}]
\sinh[\IM \gamma+3 \frac{\IM \pi k}{5}]
\ear
\bear
R_{2,2}^{2,2}(\lambda) &=& \left[ \sinh[\IM \gamma] \sinh[\IM \gamma +
\frac{\IM \pi k}{5}] + \sinh[\lambda - \frac{\IM \pi k}{5}]
\sinh[\lambda] \right] \sinh[\IM \gamma + \frac{2 \IM \pi k}{5} + \lambda]
\sinh[\IM \gamma + \frac{3 \IM \pi k}{5} + \lambda]
\nonumber \\
\ear
\bear
R_{2,3}^{2,3}(\lambda) &=& R_{3,2}^{3,2}(\lambda) = \left[2
\sinh[\IM \gamma] \sinh[\IM \gamma+\frac{2 \IM \pi k}{5}] \cosh[\frac{\IM \pi
k}{5}] + \sinh[\lambda - \frac{\IM \pi k}{5}] \sinh[\lambda] \right]
\sinh[\lambda]
\nonumber \\
& \times & \sinh[\IM \gamma + \frac{3 \IM \pi k}{5} + \lambda]
\\
R_{2,3}^{3,2}(\lambda) &=& R_{3,2}^{2,3}(\lambda) = \left[
\sinh[\IM \gamma] \sinh[\IM \gamma + \frac{2 \IM \pi k}{5}] + 2 \sinh[\lambda
- \frac{\IM \pi k}{5}] \sinh[\lambda] \cosh[\frac{\IM \pi k}{5}] \right]
\nonumber \\
&\times & \sinh[\IM \gamma + \frac{\IM \pi k}{5}] \sinh[\IM \gamma + \frac{3 I
\pi k}{5} + \lambda]
\\
R_{2,4}^{2,4}(\lambda) &=& R_{4,2}^{4,2}(\lambda) = \sinh[\lambda]
\sinh[\lambda+\frac{\IM \pi k}{5}] (\sinh[\lambda]
\sinh[\lambda-\frac{\IM \pi k}{5}]+ 2 \varepsilon_k \sinh[\IM
\gamma]
\nonumber \\
& \times & \sinh[\IM \gamma+3 \frac{\IM \pi k}{5}] \cosh[\frac{\IM
\pi k}{5}])
\\
R_{2,4}^{3,3}(\lambda) &=& R_{3,3}^{2,4}(\lambda) =
R_{4,2}^{3,3}(\lambda) = R_{3,3}^{4,2}(\lambda) = 2
\sqrt{\varepsilon_k \sinh[\IM \gamma+\frac{\IM \pi k}{5}] \sinh[\IM
\gamma+2 \frac{\IM \pi k}{5}]}
\nonumber \\
& \times & \cosh[\frac{\IM \pi k}{5}] \sinh[\lambda] (\sinh[\lambda]
\sinh[\lambda-\frac{\IM \pi k}{5}]+\sinh[\IM \gamma] \sinh[\IM
\gamma+3 \frac{\IM \pi k}{5}])
\\
R_{2,4}^{4,2}(\lambda) &=& R_{4,2}^{2,4}(\lambda) = \sinh[\IM
\gamma+\frac{\IM \pi k}{5}] \sinh[\IM \gamma+2 \frac{\IM \pi k}{5}]
(2 \varepsilon_k \sinh[\lambda] \sinh[\lambda-\frac{\IM \pi k}{5}]
\nonumber \\
& \times & \cosh[\frac{\IM \pi k}{5}]+\sinh[\IM \gamma] \sinh[\IM
\gamma+3 \frac{\IM \pi k}{5}])
\\
R_{3,3}^{3,3}(\lambda) &=& \sinh[\IM \gamma] \sinh[\IM
\gamma+\frac{I \pi k}{5}] \sinh[\IM \gamma+2 \frac{I \pi k}{5}]
\sinh[\IM \gamma+3 \frac{I \pi k}{5}]
\nonumber \\
& + & (\sinh[\lambda] \sinh[\frac{I \pi k}{5}-\lambda])^2 - 2
\cosh[\frac{I \pi k}{5}] \sinh[\lambda] \sinh[\frac{I \pi
k}{5}-\lambda]
\nonumber \\
& \times & (\sinh[\IM \gamma] \sinh[\IM \gamma+2 \frac{I \pi
k}{5}]+\sinh[\IM \gamma+\frac{I \pi k}{5}] \sinh[\IM \gamma+3
\frac{I \pi k}{5}])
\\
R_{2,5}^{2,5}(\lambda) &=& R_{5,2}^{5,2}(\lambda) = \varepsilon_k
\sinh[\IM \gamma-\lambda] \sinh[\lambda] \sinh[\lambda+\frac{\IM \pi
k}{5}] \sinh[\lambda+2 \frac{\IM \pi k}{5}]
%
\\
R_{2,5}^{3,4}(\lambda) &=& R_{3,4}^{2,5}(\lambda) =
R_{5,2}^{4,3}(\lambda) = R_{4,3}^{5,2}(\lambda) = 2 \varepsilon_k
\sinh[\IM \gamma-\lambda] \sinh[\lambda] \sinh[\lambda+\frac{\IM \pi
k}{5}]
\nonumber \\
& \times & \cosh[\frac{\IM \pi k}{5}] \sqrt{2 \cosh[\frac{2 \IM \pi
k}{5}] \sinh[\IM \gamma+\frac{\IM \pi k}{5}] \sinh[\IM \gamma+3
\frac{\IM \pi k}{5}]}
\\
R_{2,5}^{4,3}(\lambda) &=& R_{4,3}^{2,5}(\lambda) =
R_{5,2}^{3,4}(\lambda) = R_{3,4}^{5,2}(\lambda) = 2 \sinh[\IM
\gamma-\lambda] \sinh[\lambda] \sinh[\IM \gamma+2 \frac{\IM \pi
k}{5}]
\nonumber \\
& \times & \cosh[\frac{\IM \pi k}{5}] \sqrt{2 \cosh[\frac{2 \IM \pi
k}{5}] \sinh[\IM \gamma+\frac{\IM \pi k}{5}] \sinh[\IM \gamma+3
\frac{\IM \pi k}{5}]}
%
\ear
\bear
R_{2,5}^{5,2}(\lambda) &=& R_{5,2}^{2,5}(\lambda) =
\sinh[\IM \gamma-\lambda] \sinh[\IM \gamma+3 \frac{\IM \pi k}{5}]
\sinh[\IM \gamma+2 \frac{\IM \pi k}{5}] \sinh[\IM \gamma+\frac{\IM \pi k}{5}]
\ear 
\bear
R_{3,4}^{3,4}(\lambda) &=& R_{4,3}^{4,3}(\lambda) = \varepsilon_k
\sinh[\IM \gamma-\lambda] \sinh[\lambda] (\sinh[\IM \gamma+\frac{\IM
\pi k}{5}] \sinh[\IM \gamma+3 \frac{\IM \pi k}{5}]
\nonumber \\
& \times & 2 \cosh[\frac{\IM \pi k}{5}]
+ \sinh[\lambda-\frac{\IM \pi k}{5}] \sinh[\lambda])
\\
R_{3,4}^{4,3}(\lambda) &=& R_{4,3}^{3,4}(\lambda) =
\sinh[\IM \gamma-\lambda] \sinh[\IM \gamma+2 \frac{\IM \pi k}{5}]
(\sinh[\IM \gamma+\frac{\IM \pi k}{5}] \sinh[\IM \gamma+3 \frac{\IM \pi k}{5}]
\nonumber \\
& +& 2 \cosh[\frac{\IM \pi k}{5}] \sinh[\lambda-\frac{\IM \pi k}{5}]
\sinh[\lambda])
\\
R_{3,5}^{3,5}(\lambda) &=& R_{5,3}^{5,3}(\lambda) = 4 \varepsilon_k
\cosh[\frac{\IM \pi k}{5}] \cosh[\frac{2 \IM \pi k}{5}]
\sinh[\lambda] \sinh[\lambda+\frac{\IM \pi k}{5}]
\nonumber \\
& \times & \sinh[\IM \gamma-\lambda] \sinh[\IM \gamma+\frac{\IM \pi
k}{5}-\lambda]
\\
R_{3,5}^{4,4}(\lambda) &=& R_{4,4}^{3,5}(\lambda)  = 
R_{5,3}^{4,4}(\lambda) = R_{4,4}^{5,3}(\lambda)  = 
2
\cosh[\frac{\IM \pi k}{5}] \sqrt{2 \varepsilon_k \cosh[\frac{2 \IM
\pi k}{5}] \sinh[\IM \gamma+3 \frac{\IM \pi k}{5}] \sinh[\IM
\gamma+2 \frac{\IM \pi k}{5}]}
\nonumber \\
& \times & \sinh[\lambda] \sinh[\IM \gamma-\lambda] \sinh[\IM
\gamma+\frac{\IM \pi k}{5}-\lambda]
\\
R_{3,5}^{5,3}(\lambda) &=& R_{5,3}^{3,5}(\lambda) = \sinh[\IM \gamma+3
\frac{\IM \pi k}{5}] \sinh[\IM \gamma+2 \frac{\IM \pi k}{5}]
\sinh[\IM \gamma-\lambda] \sinh[\IM \gamma+\frac{\IM \pi k}{5}-\lambda]
\\
R_{4,4}^{4,4}(\lambda) &=& (\sinh[\IM \gamma+2 \frac{\IM \pi k}{5}]
\sinh[\IM \gamma+3 \frac{\IM \pi k}{5}] + 4 \varepsilon_k
\cosh[\frac{\IM \pi k}{5}] \cosh[\frac{2 \IM \pi k}{5}]
\nonumber \\
& \times & \sinh[\lambda-\frac{\IM \pi k}{5}] \sinh[\lambda])
\sinh[\IM \gamma-\lambda] \sinh[\IM \gamma+\frac{\IM \pi k}{5}-\lambda]
\\
R_{4,5}^{4,5}(\lambda) &=& R_{5,4}^{5,4}(\lambda) = \varepsilon_k
\sinh[\lambda] \sinh[\IM \gamma-\lambda] \sinh[\IM \gamma+\frac{\IM
\pi k}{5}-\lambda] \sinh[\IM \gamma+2 \frac{\IM \pi k}{5}-\lambda]
\ear
\bear
R_{4,5}^{5,4}(\lambda) &=& R_{5,4}^{4,5}(\lambda) = \sinh[\IM \gamma+3
\frac{\IM \pi k}{5}] \sinh[\IM \gamma-\lambda] \sinh[\IM \gamma+\frac{\IM \pi
k}{5}-\lambda] \sinh[\IM \gamma+2 \frac{\IM \pi k}{5}-\lambda]
\nonumber \\
\ear
\bear
R_{5,5}^{5,5}(\lambda) &=& \sinh[\IM \gamma - \lambda] \sinh[\IM
\gamma + \frac{\IM \pi k}{5} - \lambda] \sinh[\IM \gamma + \frac{2
\IM \pi k}{5} - \lambda] \sinh[\IM \gamma + \frac{3 \IM \pi k}{5} -
\lambda] 
\ear 
where the possible values of $k=1,2,3,4$.

\section*{Acknowledgments}
The authors thank the Brazilian Research Agencies FAPESP and CNPq for financial support.


\addcontentsline{toc}{section}{References}
\begin{thebibliography}{}
\bibitem{QG} V.G. Drinfeld, {\em Sov.Math.Doke. 32 (1985) 254};
M.Jimbo, {\em Commun.Math.Phys. 102 (1986) 537.}
\bibitem{BA} R.J. Baxter, {``Exactly Solved Models in Statistical Mechanics''}, Academic Press, New York, 1982.
\bibitem{QG1} M. Jimbo, {\em Lett.Math.Phys. 10 (1985) 63}
\bibitem{SO} A.B. Zamolodchikov and V.A. Fatteev, {\em Sov.J.Nucl.Phys 32 (1980)
298}; K. Sogo, Y. Akutsu and J. Abe, {\em Prog.Theor.Phys. 70
(1983) 730}
\bibitem{FA} L.D. Faddeev, V.O. Tarasov and L.A. Takhtajan, {\em Theor.Math.Phys. 57 (1983) 1059}
\bibitem{RES} A.N. Kirillov and N.Y. Reshetikhin, {\em J.Sov.Math. 23 (1983) 2435; J. Phys.A:Math.Gen. 20 (1987) 1565}
\bibitem{QG2} C. De Concini and V.G. Kac, {\em Prog.Math.Birkhauser 92 (1990)
471}; G. Lusztig, {\em Geom.Dedicata 35 (1990) 89}
\bibitem{DEG} T. Deguchi and Y. Akutsu, {\em J.Phys.Soc.Jpn. 60 (1991) 4051; Phys.Rev.Lett. 67 (1991) 777}
\bibitem{CO} M. Couture, {\em J.Phys.A: Math.Gen. 24 (1991) L 103}
\bibitem{SIE} C. Gomez, M. Ruiz-Altaba and G. Siera, {\em Phys.Lett.B 265 (1991)
95}; C. Gomez and G. Sierra, {\em Nucl.Phys.B 373 (1992) 761}
\bibitem{DEG1} T. Deguchi and Y. Akutsu, {\em Mod.Phys.Lett.A 7 (1992) 767; J.Phys.Soc.Jpn. 62 (1993) 19}
\bibitem{SIE1} A. Berkovich, C. Gomez and G. Sierra, {\em Int.J.Mod.Phys.B 61 (1992) 1939; J.Phys.A: Math.Gen. 26 (1993) L 45}
\bibitem{MC} M.J. Martins and C.S. Melo, {\em Nucl.Phys.B 820 (2009) 620}
\bibitem{CARA} O. Foda, M. Wheeler and M. Zuparic, {\em J.Stat.Mech. (2007) P10016};
A. Caradoc, O. Foda, M. Wheeler and M. Zuparic, {\em J.Stat.Mech. (2007) P03010}
\bibitem{CGM} C.S. Melo, G.A.P. Ribeiro and M.J. Martins, {\em Nucl.Phys.B 711 (2005) 565}
\bibitem{SK} E.K. Sklyanin, {\em J. Phys.A:Math.Gen. 21 (1988) 2375}
\bibitem{NEP1} L. Mezincescu and R.I. Nepomechie, {\em J. Phys.A: Math.Gen. 24 (1991) L 17; Int.J.Mod.Phys.A 6 (1991) 5231}
\bibitem{NEP2} L. Mezincescu and R.I. Nepomechie, {\em Int.J.Mod.Phys.A 7 (1992) 5657}
\bibitem{HAM} R. Cuerno and A. Gonzales-Ruiz, {\em J. Phys.A:Math.Gen. 26 (1993) L
605}; J.R. Links, M.D. Gould, {\em Int.J.Mod.Phys.B 10 (1956)
3461}
\bibitem{FAN} H. Fan, {\em Nucl.Phys.B 488 (1997) 405}
\bibitem{LIE} G.O. Li, K.J.Shi, R.H. Shi, {\em Nucl.Phys.B 670 (2003)
401}; G.L. Li, K.O. Shi and R.H. Yue, {\em Nucl.Phys.B 687 (2004)
220}
\bibitem{NEPX} R.I. Nepomechie, {\em Nucl.Phys.B 622 (2002) 615}, {\em J.Phys.A:Math.Gen. 37 (2004) 433} 
\bibitem{DOI} A. Doikou, {\em Nucl.Phys.B 668 (2003) 447}; {\em Phys.Letters.A 336 (2007) 556}
\bibitem{WY} W.L. Yang, R.I. Nepomechie and Y.Z. Zhang, {\em Phys.Lett.B 633 (2006) 664}; L. Frappat, R.I. Nepomechie
and E. Ragoucy, {\em J.Stat.Mech. P09008 (2007)}
\bibitem{MUR} R. Murgan, {\em JHEP 04 (2009) 076}; R. Murgan, R.I. Nepomechie and C. Shi, {\em J.Stat.Mech.
P08006 (2006)}
  
%
%
%
%
\end{thebibliography}
\end{document}